\documentclass[review]{elsarticle}

\def\squarebox#1{\hbox to #1{\hfill\vbox to #1{\vfill}}}
\renewcommand{\qed}{\hspace*{\fill}
            \vbox{\hrule\hbox{\vrule\squarebox{.667em}\vrule}\hrule}\smallskip\newline}

\newtheorem{thm}{Theorem}
\newtheorem{lem}[thm]{Lemma}
\newtheorem{cor}[thm]{Corollary}

\newproof{pf}{Proof}
\newdefinition{defi}{Definition}
\usepackage{amssymb, amsmath, enumerate}
\usepackage{graphicx}
\newcommand{\diD}{\vec{D}}
\input{epsf}

\begin{document}
\begin{frontmatter}
\title{A linear time algorithm for the next-to-shortest path problem on undirected graphs with nonnegative edge lengths}
\author[bang]{Bang Ye Wu\corref{cor1}}
\ead{bangye@cs.ccu.edu.tw}
\author[yl]{Jun-Lin Guo}
\author[yl]{Yue-Li Wang}
\address[bang]{Dept. of Computer Science and Information Engineering,
National Chung Cheng University, ChiaYi, Taiwan 621, R.O.C.}
\address[yl]{Department of Information Management, National Taiwan University of Science and Technology, Taipei, Taiwan, R.O.C.}
\cortext[cor1]{corresponding author}

\begin{abstract}
For two vertices $s$ and $t$ in a graph $G=(V,E)$, the
next-to-shortest path is an $st$-path which length is minimum
amongst all $st$-paths strictly longer than the shortest path
length. In this paper we show that, when the graph is undirected and
all edge lengths are nonnegative,  the problem can be solved in
linear time if the distances from $s$ and $t$ to all other vertices
are given.
\end{abstract}

\end{frontmatter}

\section{Introduction}
Let $G=(V,E,w)$ be an undirected graph with vertex set $V$, edge set
$E$ and edge-length function $w$. We shall use $n$ and $m$ to stand
for $|V|$ and $|E|$, respectively. For $s,t\in V$, a {\em simple
$st$-path} is a path from $s$ to $t$ without repeated vertex in the
path. In this paper, a path always means a simple path. The {\em
length of a path} is the total length of all edges in the path. An
$st$-path is a shortest $st$-path if its length is minimum amongst
all possible $st$-paths. The shortest path length from $s$ to $t$ is
denoted by $d(s,t)$ which is the length of their shortest path. A
\emph{next-to-shortest} $st$-path is an $st$-path which length is
minimum amongst those the path lengths \emph{strictly larger} than
$d(s,t)$. And the {\em next-to-shortest path problem} is to find a
next-to-shortest $st$-path for given $G$, $s$ and $t$. In this
paper, we present a linear time algorithm for solving the
next-to-shortest path problem on graphs with nonnegative edge
lengths, assuming the distances from $s$ and $t$ to all other
vertices are given.


\subsubsection*{History}
The next-to-shortest path problem was first studied by Lalgudi and
Papaefthymiou in the directed version with no restriction to
positive edge length \cite{lal97}. They showed that the problem is
intractable for path and can be efficiently solved for walk
(allowing repeated vertices). Algorithms for the problem on special
graphs were also studied \cite{bar07,mod06}. For undirected graphs
with positive edge lengths, the first polynomial algorithm was
presented in \cite{kra04} with time complexity $O(n^3m)$ time. The
time complexity has been improved several times
\cite{li06,kao11,wu10}. The currently best result is $O(m+n\log n)$
\cite{wu10}, and recently the author further improved to linear
time, assuming the distances from $s$ and $t$ to all other vertices
are given. Hence, the positive length version of the
next-to-shortest path problem can be solved with the same time
complexity as the single source shortest paths problem. On the other
hand, the problem becomes more complicated when edges of zero weight
are allowed, and there is no polynomial time algorithm for this
version before this work.

\subsubsection*{Techniques}
An edge of zero-length is called as \emph{zero-edge} and otherwise a
\emph{positive edge}. Let $D$ be the union of all shortest
$st$-paths. Let $\diD$ be the digraph obtained from $D$ by
orientating all edges toward $t$. That is, for any directed edge
(arc) in $\diD$, there is a shortest $st$-path in $G$ passing
through this edge with the same direction. Since a next-to-shortest
path either contains an edge in $E-E(D)$ or not, the problem is
divided into two subproblems: the shortest detour path problem and
the shortest zigzag path problem. The {\em shortest detour path
problem} is to find a shortest $st$-path using at least one edge not
in $E(D)$ while the {\em shortest zigzag path problem} looks for a
shortest $st$-path consisting of only edges in $E(D)$ with at least
one reverse arc of a positive length in $\diD$. Clearly, the shorter
path found from the above two subproblems is a next-to-shortest
path.

In this paper, we solve the nonnegative length version also by
solving the two subproblems individually. But there are some
difficulties to be overcome. First, the digraph $\diD$ is not so
easy to construct as in the positive length version. Secondly,
$\diD$ is no more a DAG (directed acyclic graph) as in the positive
length version, and therefore some properties in \cite{wu10} cannot
be used. Instead $\diD$, we solve the two subproblems based on a
relaxed digraph $D^+$ of $\diD$, in which all zero edges are
regarded as bidirectional. The method to solve the shortest detour
path subproblem is similar to the previous one for the positive
length version, but a special care is taken into consideration for
the zero-edges and the proofs are non-trivial and different from the previous ones.

The shortest zigzag path subproblem is relatively more complicated.
To solve this subproblem efficiently, the most important thing is to
determine for a pair of vertices $(x,y)$ if there exists a simple
$st$-path using a path from $x$ to $y$ as a backward subpath. The
previous paper \cite{wu10} showed a necessary and sufficient
condition for the positive length version, but this condition no
more holds when there are zero-edges. To overcome this difficulty,
we use immediate dominators developed in the area of flow analysis.
In addition, we define zero-component in $D^+$, which are basically
connected components of the subgraph induced by the zero-edges but
any vertex and its dominators are divided into different components.
By shrinking zero-components and orientating the remaining
zero-edges, we construct an auxiliary DAG. With the help of the
auxiliary DAG, we categorize a shortest zigzag path into four types
and derive necessary and sufficient conditions individually.

The main result of this paper is the following theorem, and its
proof is given by Theorems~\ref{thm:back} and \ref{thm:out} in
Sections~3 and 4, respectively.
\begin{thm}\label{thm:main}
A next-to-shortest $st$-path of an undirected graph with nonnegative
edge lengths can be found in linear time if the distances from $s$
and $t$ to all other vertices are given.
\end{thm}

\subsubsection*{Paper organization}
The paper is divided as follows. In Section 2, the preliminaries are
presented. In addition to the notation used in this paper, in the
preliminaries, we introduce dominators, a method of constructing
$D^+$, and zero-components. Also we show some basic properties in
this section. In Sections 3 and 4, we discuss the shortest zigzag,
and detour, path problems, respectively. And finally concluding
remarks are given in Section 5.

\section{Preliminaries}
\subsection{Notation and some properties}
Throughout this paper, we shall assume that $G$ is the input graph
and $(s,t)$ is the pair of vertices for which a next-to-shortest
path is asked. Furthermore, $G$ is simple, connected and undirected,
and all edge lengths are nonnegative integers.

For a graph $H$, $V(H)$ and $E(H)$ denote its vertex and edge sets,
respectively. For simplicity, sometimes we abuse the notation of a
subgraph for its vertex set when there is no confusion from the
context. A {\em $uv$-path} is a path from $u$ to $v$. For vertices
$u$ and $v$ on path $P$, let $P[u,v]$ denote the subpath from $u$ to
$v$. We shall use ``a $uv$-path'' and a path $P[u,v]$ alternatively.
For a path $P$, we use $\bar{P}$ to denote the reverse path of $P$.
For paths $P[v_1,v_2]$ and $Q[v_2,v_3]$, $P\circ Q$ denotes the path
obtained by concatenating these two paths. Note that, even for an
undirected path, we use $P[u,v]$ to specify the direction from $u$
to $v$. For example, by ``the first vertex $x$ of $P[u,v]$
satisfying some property'', we mean that $x$ is the first vertex
satisfying the property when we go from $u$ to $v$ along path $P$.
Two paths are {\em internally disjoint} if they have no common
vertex except their endpoints. For a path $P$, let $w(P)=\sum_{e\in
E(P)}w(e)$ denote the length of the path. Let $d(u,v)$ denote the
shortest path length from $u$ to $v$ in $G$, which is also called
the {\em distance} from $u$ to $v$. For convenience, let
$d_s(v)=d(s,v)$ and $d_t(v)=d(v,t)$.

To show the time complexities more precisely, we shall assume the
distances from $s$ and $t$ to all other vertices are given. These
distances can be found by solving the single source shortest paths
(SSSP) problem. For general undirected and nonnegative edge length
graphs (the most general setting of the problem discussed in this
paper), the SSSP problem can be solved in $O(m+n\log n)$ time
\cite{cor01,fred87}, and more efficient algorithms exist for special
graphs or graphs with restrictions on edge lengths. A shortest path
tree rooted at $s$ can also be constructed in linear time if the
distances from $s$ to all others are given.

\subsection{$D$ and $D^+$}

Let $D^+$ be the digraph obtained from $D$ by orientating all
positive edges toward $t$. That is, we treat all zero-edges as
bidirectional even though only one direction of some of them can be
used to form a shortest $st$-path. Our algorithm for finding a
shortest zigzag path works on $D^+$ for the sake of efficiency.

To construct $D^+$, we have to construct $D$ first. In the
following, we show how to construct $D$ and $D^+$ in linear time.
Clearly, for $v\in V(D)$, $d_s(v)+d_t(v)=d(s,t)$ always holds.
Unfortunately, the condition that $d_s(v)+d_t(v)=d(s,t)$ is not a
necessary and sufficient condition to determine the set of vertices
in $V(D)$ when there are zero-edges. The reason is described as
follows. Let $D'$ be the subgraph of $G$ with
$V(D')=\{v|d_s(v)+d_t(v)=d(s,t)\}$ and $E(D')=
\{(u,v)|d_s(v)=d_s(u)+w(u,v)\}$ for $u,v\in V(D')$. A vertex is a
{\em non-$st$-cut} if it is a cut vertex and its removal does not
separate $s$ and $t$. For a non-$st$-cut $x$, a connected component
$K$ of $D'-x$ is called a {\em knob} if $s,t\notin V(K)$. Since $x$
is a cut vertex, any $st$-path passing through a vertex in $K$
repeats at $x$ and cannot be simple. Furthermore, for any vertex $v$
in $K$, since $d_s(v)+d_t(v)=d(s,t)$, it must be connected to $x$ by
a path of zero-length.

\begin{lem}\label{D+}
A vertex $v$ is in $V(D)$ iff $v\in V(D')$ is not in any knob.
\end{lem}
\begin{pf}
By definition, $v\in V(D)$ implies $v\in V(D')$. Furthermore, $v$
cannot be in any knob since there is no simple $st$-path in $D'$
passing through any vertex in a knob.

Now, we prove the other direction. For any vertex $v\in
V(D')-\{s,t\}$, consider the digraph $D''$ obtained from $D'$ by
reversing the direction of all positive edges $(x,y)$ with
$d_s(y)>d_s(v)$. Also we add a new vertex $s_0$ as well as two arcs
$(s_0,s)$ and $(s_0,t)$. Then there exists a shortest $st$-path
passing $v$ in $D'$ iff there are two disjoint paths from $s_0$ to
$v$ in $D''$, or equivalently there is no non-$st$-cut. Obviously
any vertex $u$ with $d_s(u)\neq d_s(v)$ cannot be an $s_0v$-cut in
$D''$, and there exists such a cut node iff $v$ is in a knob.
\qed\end{pf}

\begin{lem}
$D^+$ can be constructed in linear time if $d_s(v)$ and $d_t(v)$ are
given for all $v$.
\end{lem}
\begin{pf}
First we construct $D'$ in linear time. By using depth-first search
starting from $s$, all cut vertices can be found in linear time.
According to the order of found cut vertices, all knobs can be
detected by checking the components after removing the cut
vertices.
\qed\end{pf}

\subsection{Dominators in $D^+$}

We shall use the term ``immediate dominators'' defined in
\cite{als99}. A vertex $v\in V(D^+)$ is an {\em $s$-dominator} of
another vertex $u$ iff all paths from $s$ to $u$ contain $v$. An
$s$-dominator $v$ of $u$ is an \emph{$s$-immediate-dominator} of
$u$, denoted by $I_s(u)$, if it is the one closest to $u$, i.e., any
other $s$-dominator of $u$ is an $s$-dominator of $I_s(u)$. In
$D^+$, any vertex has a unique $s$-immediate-dominator. The
$t$-dominator is defined symmetrically, i.e., $v$ is a {\em
$t$-dominator} of $u$ iff any $ut$-path contains $v$, and $I_t(u)$
stands for the $t$-dominator closest to $u$. Note that $s$ is an
$s$-dominator and $t$ is a $t$-dominator of any other vertex in
$D^+$.

Finding immediate dominators is one of the most fundamental problems
in the area of global flow analysis and program optimization. The
first algorithm for the problem was proposed in 1969 by Lowry and
Medlock \cite{lor69}, and then had been improved several times
\cite{har85,len79,pur72,tar74}. A linear time algorithm for finding
the immediate dominator for each vertex was given in \cite{als99}.

\subsection{Zero-components}
\mbox{}

\begin{defi} A path $P$ is a 0-path if all
edges in $P$ are zero-edges. A 0-path $P[u,v]$ is a
$0^*$-path if $P[u,v]$ does not contain any vertex in
$\{I_s(u),I_s(v),I_t(u),I_t(v)\}$. A {\em zero-component} is the subgraph
of $D^+$ induced by a maximal set of vertices in which every two
vertices are connected by a $0^*$-path. The zero-component which $v$
belongs to is denoted by $Z(v)$.
\end{defi}

A zero-component may contain only one vertex but no edge. All the
zero-components partition $V(D)$ into equivalence classes, i.e.,
$v_1\in Z(v_2)$ iff $v_2\in Z(v_1)$. We shall show how to find all
zero-components of $D^+$ in linear time.

\begin{lem}\label{zero-d}
If $v'\in Z(v)$, then $I_s(v')=I_s(v)$ and $I_t(v')=I_t(v)$.
\end{lem}
\begin{pf}
If $I_s(v')$ is not an $s$-dominator of $v$, there is an $sv$-path
$Q_1$ avoiding $I_s(v')$. Since $v$ and $v'$ are in the same
zero-component, there is a 0-path $Q_2[v,v']$ in $Z(v)$ avoiding
$I_s(v')$. Thus, $Q_1\circ Q_2$, possibly taking a short-cut if the
path is non-simple, is a path from $s$ to $v'$ avoiding $I_s(v')$, a
contradiction. Therefore $I_s(v')$ is an $s$-dominator of $v$.
Similarly we can show that $I_s(v)$ is also an $s$-dominator of
$v'$. Consequently $I_s(v')$ and $I_s(v)$ dominate each other, and
thus they are the same vertex. The result $I_t(v')=I_t(v)$ can be
shown similarly.
\qed\end{pf}

An \emph{$s$-dominator tree} \cite{als99} of $D^+$ is a tree $T$
with root $s$ and vertex set $V(D^+)$. A vertex $u$ is a child of
$v$ in $T$ iff $v=I_s(u)$.
\begin{lem}
The subgraph of $D^+$ induced by the edge set
$E_0-E(T_s^d)-E(T_t^d)$ is the union of all zero-components, where
$E_0$ is the zero-edges set, and $E(T_s^d)$ and $E(T_t^d)$ are the
edge sets of $s$- and $t$-dominator trees of $D^+$, respectively.
\end{lem}
\begin{pf}
Since no positive edge is in any zero-component, we only need to
consider the zero-edges $E_0$. For any vertex $v$, if $(u,I_s(v))$
is the last edge of a path from $v$ to $I_s(v)$, then $u$ is a child
of $I_s(v)$ in the $s$-dominator tree. After removing $E(T_s^d)$ and
$E(T_t^d)$, there is no path from any vertex to its $s$- or
$t$-dominator. Therefore, by definition, the induced subgraph is the
union of all zero-components.
\qed\end{pf}

Since a dominator tree can be constructed in linear time
\cite{als99}, the next corollary follows directly from the above
lemma.

\begin{cor}
All zero-components of $D^+$ can be found in linear time.
\end{cor}

\subsection{Outward and backward subpaths}

A positive edge $(u,v)\in E$ is a reverse positive edge if $(v,u)\in
E(D^+)$. It implies that $(u,v)\notin E(D^+)$ since any positive
edge in $D^+$ is unidirectional.

\begin{defi}
A {\em backward subpath} of a path in $G$ is a path consisting of at
least one reverse positive edge and possibly some zero-edges. A
semi-path in $D^+$ with at least one backward subpath is called a
{\em zigzag path}. Two backward subpaths in a zigzag path are
consecutive if there are separated by a sequence of non-reverse
positive edges and zero edges; otherwise, they form a longer
backward subpath indeed.\footnotemark[4]
\footnotetext[4]{Another way to define a
backward subpath is a {\em maximal} subpath consisting of at least
one reverse positive edge and possibly some zero-edges. The
difference is that, by our definition, there may be some zero-edges
preceding or succeeding a backward subpath. Our definition is for
the sake of simplifying some proofs.}
\end{defi}

By definition, a zigzag path is a semi-path in $D^+$. For
simplicity, we shall use ``path'' instead of ``semi-path'' in the
following.

\begin{defi}
An {\em outward subpath} of an $st$-path in $G$ is a path consisting
of edges in $E-E(D)$. The both endpoints of an outward subpath are
in $V(D)$ and all its internal vertices, if any, are not in $V(D)$.
An $st$-path is called a {\em detour path} if it contains at least
one outward subpath.
\end{defi}

The {\em shortest detour path problem} is to find a shortest detour
$st$-path while the {\em shortest zigzag path problem} looks for a
shortest zigzag $st$-path consisting of only edges in $E(D)$. Since
a next-to-shortest path either contains an edge in $E-E(D)$ or not,
the shorter path found from the above two subproblems is a
next-to-shortest path. Since $s$ and $t$ are fixed throughout this
paper, we shall simply use ``zigzag path'' and ``detour path''
instead of ``zigzag $st$-path'' and ``detour $st$-path'',
respectively.

When the edge lengths are all positive, the following result was
shown in \cite{kao11}, and it is also the basis of the algorithms in
this paper. In remaining paragraphs of this subsection, we show
Theorem~\ref{backout} by Lemmas~\ref{oneback} and \ref{oneout}.

\begin{thm}\label{backout}
A shortest zigzag path contains exactly one backward subpath. A
shortest detour path contains exactly one outward subpath and no
backward subpath.
\end{thm}

\begin{figure}[t]
\begin{center}
\includegraphics[scale=1.2]{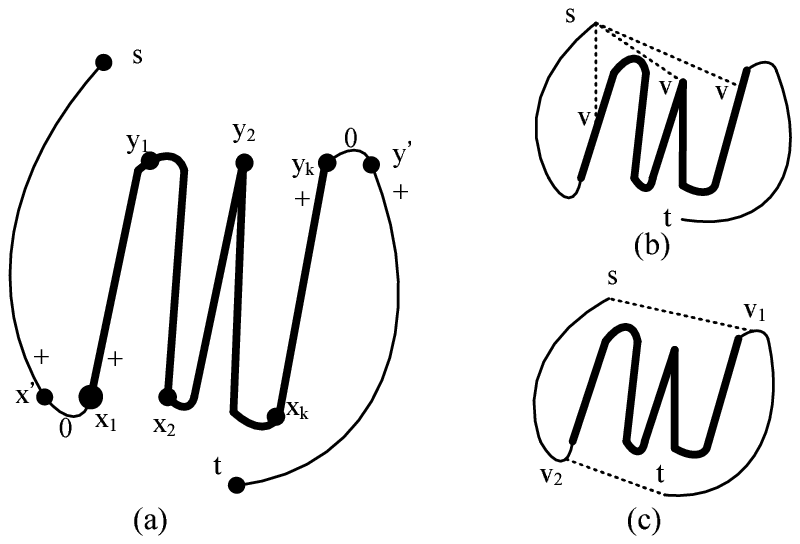}
\caption{Illustrations for {\rm Lemma~\ref{oneback}}. {\rm (a)} A path $P$ with more than one backward subpath.
The bold line is $Q$; {\rm (b)} Case 1; {\rm (c)} Case 3.}
\label{f1back}
\end{center}
\end{figure}

\begin{lem}\label{oneback}
A shortest zigzag path contains exactly one backward subpath.
\end{lem}
\begin{pf}
Suppose by contradiction that $P$ is a shortest zigzag path in $D^+$
with more than one backward subpath. Let $P[x_i, y_i]$, for
$1\leqslant i\leqslant k$, be the consecutive backward subpaths in
$P$ and $Q=P[x_1, y_k]$ where $d_s(x_i)> d_s(y_i)$ and $k\geqslant
2$. We may assume that the first and the last edges of $Q$ are
positive edges (otherwise move $x_1$ forth or $y_k$ back
accordingly). Let $x'$ be the first vertex on $P$ such that
$w(P[x',x_1])=0$ and $y'$ the last vertex such that $w(P[y_k,y'])=0$
(see Fig.~\ref{f1back}.(a)). We divide into three cases, and in
either case we show that there exists a shorter zigzag path $P'$.

\begin{itemize}
\item There is a path $P_1$ from $s$ to an internal vertex $v$ of $Q$ such that $P_1$ is disjoint to $P[y_k,y']$. Then $P'=P_1\circ P[v,t]$ is a zigzag path. Since $P_1$ is a short-cut of $P[s,v]$, $P'$ is shorter than $P$ (see Fig.~\ref{f1back}.(b)).
\item There is a path $P_2$ from an internal vertex $v$ of $Q$ to $t$ such that $P_2$ is disjoint to $P[x',x_1]$. Similarly, $P'=P[s,v]\circ P_2$ is a zigzag path shorter than $P$.
\item Otherwise, since the first case does not hold, there exists a path $P_1$ from $s$ to a vertex $v_1$ on $P[y_k,y']$, which is internally disjoint to $Q$. Furthermore, $d_s(y')\leqslant d_s(y_1)<d_s(x')$. Similarly, there exists a path $P_2$ from a vertex $v_2$ on $P[x',x_1]$ to $t$, which is internally disjoint to $Q$. And $d_t(x')\leqslant d_t(x_k)<d_t(y')$. Then the path $P'=P_1\circ \bar{P}[v_1,v_2]\circ P_2$ is a zigzag path. Clearly $w(P')=w(P)-d_s(x')-d_t(y')+d_s(y')+d_t(x')<w(P)$ (see Fig.~\ref{f1back}.(c)).
\end{itemize}
\qed\end{pf}

\begin{lem}\label{2paths}
For any two vertices $x$ and $y$ in $V(D^+)-\{s,t\}$, there exist an
$sx$-path and a $yt$-path; or an $sy$-path and an $xt$-path; which
are disjoint.
\end{lem}
\begin{pf}
The result is trivial for the case $x\notin Z(y)$. We only need to
show the case $x\in Z(y)$. To show the lemma for this case, we
construct an auxiliary directed graph from $D^+$ by adding a new
vertex $v$ and two bidirectional edges $(v,x)$ and $(v,y)$. Since
there is no  non-$st$-cut, similar to Menger's theorem, there is an
$st$-path passing through $v$ in the auxiliary graph, and the
desired two paths exist.
\qed\end{pf}

\begin{lem}\label{oneout}
A shortest detour path contains exactly one outward subpath and no
backward subpath.
\end{lem}
\begin{pf}
Let $P$ be a shortest detour path, in which $P[x,y]$ is an outward
subpath. We shall show that if $P$ had another outward subpath or
backward subpath in addition to $P[x,y]$, we could construct a
detour path $P'$ shorter than $P$.

By Lemma~\ref{2paths}, there exist an $sx$-path and a $yt$-path; or
an $sy$-path and an $xt$-path in $D^+$ which are disjoint. In either
case that the two paths exist, we can concatenate the two paths with
$P[x,y]$ (or its reverse) to form a simple $st$-path. It is clear
that the shorter detour path in the two cases is a shortest detour
path $P'$ containing $P[x,y]$.
\qed\end{pf}

\section{Shortest zigzag path}

\subsection{Basic properties}
By Theorem~\ref{oneback}, a shortest zigzag path has the form
$P^*=P_1[s,x]\circ \bar{P}_2[x,y]\circ P_3[y,t]$, in which $P_i$ are
paths in $D^+$. Since $P^*$ is required to be simple, the three
subpaths must be simple and disjoint except at the two joint
vertices. Therefore our goal is to find $x,y\in V(D)$ minimizing
\begin{equation}
d(s,x)+d(x,y)+d(y,t)=d(s,t)+2d(y,x)
\end{equation}
subject to that there exists a simple path $P_1[s,x]\circ
\bar{P}_2[x,y]\circ P_3[y,t]$ in $D^+$. Since $d(s,t)$ is fixed for
a given graph $G$, the objective is to find the minimum of $d(y,x)$.
If $x$ and $y$ satisfy the constraint, we say ``the pair $(x,y)$ is
{\em valid}'' and ``$y$ is valid for $x$''. A valid pair $(x,y)$
with minimum $d(y,x)$ is an \emph{optimal backward pair}, or simply
{\em optimal pair}, and the corresponding backward subpath is an
{\em optimal backward subpath}. The shortest zigzag path problem is
equivalent to finding an optimal pair.

The auxiliary simple digraph $\mathcal{Z}$ is obtained from $D^+$ by
shrinking every zero-component and orientating all the remaining
zero-edges toward $t$. By the definition of zero-component, if
$(u,v)$ is a zero-edge in $\mathcal{Z}$, $u=I_s(v)$ or $v=I_t(u)$.
Therefore the orientation can be easily done. For $Z(v)$ in $D^+$,
let $z_v$ denote its corresponding vertex in $\mathcal{Z}$. For
simplicity, since $s$ and $t$ themselves must be zero-components,
the corresponding vertices in $\mathcal{Z}$ are also denoted by $s$
and $t$, respectively. For a vertex $z_v$, $I_s(z_v)$ and $I_t(z_v)$
are again the immediate $s$- and $t$-dominators (but in
$\mathcal{Z}$). A simple path in $D^+$ corresponds to a simple path
in $\mathcal{Z}$ since, without backward subpath, a path cannot
enter a zero-component twice.

\begin{defi}
We define a binary relation on pairs of vertices in $V(D^+)$:
$u\prec v$ or equivalently $v\succ u$ iff $z_u\neq z_v$ and there
exists a path from $z_u$ to $z_v$ in $\mathcal{Z}$. Let $C_s(u)=\{v|
I_s(u)\prec v \wedge v\prec u\}$ and $C_t(u)=\{v| v\prec I_t(u)
\wedge u\prec v\}$.
\end{defi}

\begin{defi}
The predicate $\beta_1(x,y)$ is true iff $y\in C_s(x)$ and $x\in
C_t(y)$.
\end{defi}

\begin{lem}\label{beta1a}
If $\beta_1(x,y)$ is true, $d_s(y)<d_s(x)$.
\end{lem}
\begin{pf}
By definition, $y\in C_s(x)$, and therefore $d_s(y)\leqslant d_s(x)$
and $y\notin Z(x)$. If $d_s(y)=d_s(x)$, they are connected by a
0-path but not a $0^*$-path, i.e., a path containing a vertex in
$\{I_s(x),I_s(y),I_t(x),I_t(y)\}$. Since $y\prec x$, a $yx$-path
contains neither $I_s(y)$ nor $I_t(x)$. Since $y\in C_s(x)$ and
$x\in C_t(y)$, $I_s(x)\prec y\prec x\prec I_t(y)$, which implies
that any $yx$-path in $D^+$ contains neither $I_s(x)$ nor $I_t(y)$,
a contradiction.
\qed\end{pf}

The notation defined on $D^+$ will also be used for $\mathcal{Z}$.
We do not distinguish between them since there will be no confusion
from the context. The next two lemmas appeared in \cite{wu10} for
positive length version, and it is easy to see it also holds for
nonnegative length version. The next lemma show a necessary
condition for the validity of a pair.
\begin{lem}\label{back_nec}
If $(x,y)$ is valid, then $\beta_1(x,y)$ is true.
\end{lem}
\begin{pf}
By definition, $y\prec x$. If $I_s(x)\not\prec y$, by the definition
of immediate dominator, any $sx$-path and $yx$-path contain $I_s(x)$
simultaneously and cannot be disjoint. Therefore we have
$I_s(x)\prec y$, and then $y\in C_s(x)$ by definition. The relation
$x\in C_t(y)$ can be shown similarly.
\qed\end{pf}

\begin{lem}\label{close}
If $y\in C_s(x)$, there are two paths from $s$ and $y$,
respectively, to $x$, which are  disjoint except at $x$.
\end{lem}
\begin{pf}
Let $p=I_s(x)$ and $R$ be any $sp$-path. By the definition of
immediate dominator, removing any vertex in $C_s(x)$ cannot separate
$p$ and $x$ and therefore there are two internally disjoint
$px$-paths, say $P_1$ and $P_2$. If $y$ is on one of them, say
$P_2$, we have done since $R\circ P_1$ and $P_2[y,x]$ are the
desired paths. Otherwise, let $P_3$ be any $yx$-path and $v$ be the
first vertex on $P_3$ and also in $V(P_1)\cup V(P_2)$. W.l.o.g. let
$v\in V(P_1)$. Then, the path $P_3[y,v]\circ P_1[v,x]$ is a
$yx$-path disjoint to $R\circ P_2$.
\qed\end{pf}

\begin{lem}\label{redu_y}
If $\beta_1(x,y)$ is true and there exists a path $P$ from $y$ to
$t$ avoiding $Z(x)$, then there exists a vertex valid for $x$.
Furthermore if $y^*$ satisfies the above condition with minimum
$d(y^*,x)$, then there exists a vertex $v$ such that $(x,v)$ is
valid and $d(v,x)=d(y^*,x)$. The same result also holds for the case
that $\beta_1(x,y)$ is true and there exists a path from $s$ to $x$
avoiding $Z(y)$.
\end{lem}
\begin{pf}
We show the first result and the second one can be shown similarly.
Let $v$ be the last vertex of $P$ in $C_s(x)$. Since $\beta_1(x,v)$
is also true, we have that $d_s(y)\leqslant d_s(v)<d_s(x)$ by
Lemma~\ref{beta1a}. By Lemma~\ref{close}, there are a path
$P_1[s,x]$ and a path $P_2[v,x]$ which are internally disjoint.
Then, the path $P_1\circ \bar{P}_2\circ P[v,t]$ is a zigzag path and
therefore $(x,v)$ is a valid pair.

Since $v$ is also a vertex satisfying the condition, we have
$d(v,x)=d(y^*,x)$, otherwise $v$ contradicts the minimality of
$y^*$.
\qed\end{pf}

\subsubsection*{Types of optimal backward pairs}

By the definition of zero-component, there exists an $sx$-path
avoiding $Z(y)$ iff $z_y$ is not an $s$-dominator of $z_x$.
Similarly, there exists a $yt$-path avoiding $Z(x)$ iff $z_x$ is not
a $t$-dominator of $z_y$. Therefore, all the valid pairs $(x,y)$ can
be categorized into the following four types, and the best of the
four types, if any, is an optimal pair.
\begin{itemize}
\item Type I: $z_y$ is not an $s$-dominator of $z_x$ and $z_x$ is not a $t$-dominator of $z_y$.
\item Type II: $z_y$ is an $s$-dominator of $z_x$ and $z_x$ is a $t$-dominator of $z_y$.
\item Type III: $z_y$ is an $s$-dominator of $z_x$ and $z_x$ is not a $t$-dominator of $z_y$.
\item Type IV: $z_y$ is not an $s$-dominator of $z_x$ and $z_x$ is a $t$-dominator of $z_y$.
\end{itemize}

In the following subsections, we shall derive linear time algorithms
for each of the types. The next theorem concludes the result of this
section, and its proof is given by Lemmas \ref{type1t}, \ref{type2t}
and \ref{type34t} in the following subsections.

\begin{thm}\label{thm:back}
Suppose that $d_s(v)$ and $d_t(v)$ are given for all vertices $v$. A
shortest zigzag path can be found in linear time.
\end{thm}

\subsection{Type I}

By definition, $\mathcal{Z}$ is a DAG. If $(z_u,z_v)$ is a zero-edge
in $\mathcal{Z}$, then  $z_u=I_s(z_v)$ or $z_v=I_t(z_u)$. By this
property, all the properties and the algorithm derived for a
shortest zigzag path in \cite{wu10} also hold for $\mathcal{Z}$.
\begin{lem}\label{type1t}
Suppose that $d_s(v)$ and $d_t(v)$ are given for all vertices $v$. A
shortest zigzag path of type I can be found in linear time.
\end{lem}
\begin{pf}
For $(x,y)$ such that $\beta_1(x,y)$ is true, by definition, the
pair $(x,y)$ is valid of type I iff $(z_x,z_y)$ is valid in
$\mathcal{Z}$. Therefore, a shortest zigzag path of type I in $D^+$
can be found by solving the shortest zigzag path problem in
$\mathcal{Z}$. By the result of \cite{wu10}, it can be done in
linear time.\footnotemark[4]
\qed\end{pf}
\footnotetext[4]{The problem is named {\em the optimal
backward problem} in \cite{wu10}.}

\subsection{Type II}

For a shortest zigzag path of type II, the corresponding path in
$\mathcal{Z}$ repeats at both $z_x$ and $z_y$. The next lemma is not
only for type II.

\begin{figure}[t]
\begin{center}
\includegraphics[scale=1.2]{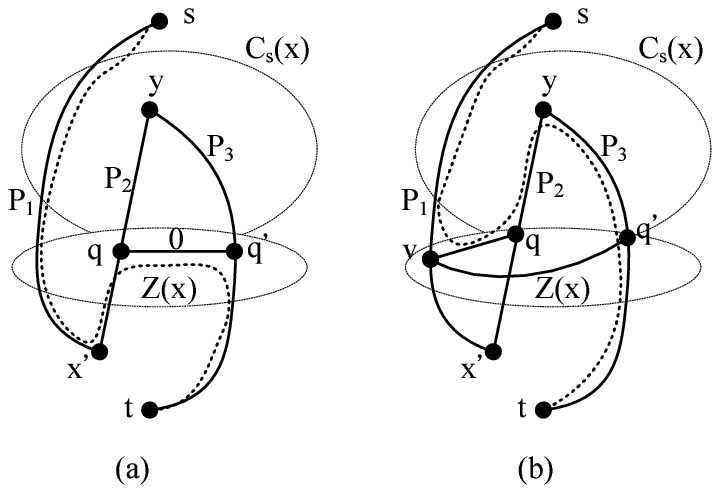}
\caption{Illustrations for {\rm Lemma~\ref{redu_cx}}. The dotted-line depicts a valid zigzag path. {\rm (a)} $V(Q)\cap V(P)=\emptyset$ and $(x',q)$ is valid; {\rm (b)} $Q$ intersects $P_1$ and $(v,y)$ is valid.}
\label{f2back}
\end{center}
\end{figure}
\begin{lem}\label{redu_cx}
For any $y\in C_s(x)$, if $y$ is valid for some $x'\succ x$, then
there exists some $v\in Z(x)$ such that $(v,y)$ or $(x',v)$ is
valid.
\end{lem}
\begin{pf}
Since $y$ is valid for $x'$, there exists a path $P=P_1[s,x']\circ
\bar{P}_2[x',y]\circ P_3[y,t]$. By Lemma~\ref{back_nec},
$\beta_1(x',y)$ is true and $x'\prec I_t(y)$. Since $y\in C_s(x)$
and $x\prec x'\prec I_t(y)$, $\beta_1(x,y)$ is also true.

If $\bar{P}_2$ or $P_3$ does not pass any vertex in $Z(x)$, by
Lemma~\ref{redu_y}, $(x,y)$ is valid and the proof is complete.
Otherwise both the two subpaths pass vertices in $Z(x)$, and
therefore, in $Z(x)$ we can find $q$ and $q'$ on $\bar{P}_2$ and
$P_3$, respectively, as well as a 0-path $Q[q,q']$ which is
internally disjoint to $\bar{P}_2$ and $P_3$. Since $y\in C_s(x')$
and $y\prec x\prec x'$, $x\in C_s(x')$. Since $P_3$ is a path
passing $Z(x)$ and avoiding $x'$, $x'\in C_t(x)$. Therefore
$d_s(x')>d_s(x)$ by Lemma~\ref{beta1a}.

If $Q$ is disjoint to $P_1$, the path $P=P_1[s,x']\circ
\bar{P}_2[x',q]\circ Q\circ P_3[q',t]$ is a simple path with a
backward subpath from $x'$ to $q$. That is, $q\in Z(x)$ and $(x',q)$
is valid (Fig.~\ref{f2back}.(a)). Otherwise $Q$ intersects $P_1$.
Let $v$ be the intersection vertex closest to $q$ on $Q$. Then, the
path $P=P_1[s,v]\circ Q[v,q]\circ\bar{P}_2[q,y]\circ P_3[y,t]$ is a
simple path with a backward subpath from $v$ to $y$. That is, $v\in
Z(x)$ and $(v,y)$ is valid (Fig.~\ref{f2back}.(b)).
\qed\end{pf}

\begin{lem}\label{2p1to2}
For any two vertices $u$ and $v$ such that $I_s(u)=I_s(v)=p$, there
exist two internally disjoint paths from $p$ to $u$ and $v$,
respectively.
\end{lem}
\begin{pf}
By definition, there exists no cut vertex whose removal separates
$I_s(v)$ from $u$ or $v$. By Menger's theorem, such two disjoint
paths exist.
\qed\end{pf}

By definition, if $(x,y)$ is valid for type II, $z_x$ is a
$t$-dominator of $z_y$ and $z_y$ is an $s$-dominator of $z_x$. We
show a stronger condition in the next lemma.

\begin{lem}\label{type2b}
If $(x,y)$ is an optimal pair of type II, $z_x=I_t(z_y)$ and
$z_y=I_s(z_x)$.
\end{lem}
\begin{pf}
Suppose that $P=P_1\circ \bar{P}_2\circ P_3$ is a shortest zigzag
path of type II, in which $\bar{P}_2$ is the backward subpath from
$x$ to $y$. If $z_y\neq I_s(z_x)=z_v$, both $P_1$ and $P_2$ contain
a vertex in $Z(v)$. We shall show that $y\in C_s(v)$, and then by
Lemma~\ref{redu_cx}, $(x,y)$ is not optimal. The result
$z_x=I_t(z_y)$ can be handled similarly.

Let $P_1'[y_1,v_1]$ and $P_2'[y_2,v_2]$ be subpaths of $P_1$ and
$P_2$, respectively, such that $\{y_1,y_2\}\subset Z(y)$,
$\{v_1,v_2\}\subset Z(v)$, and no internal vertex of them is in
$Z(y)\cup Z(v)$. We can find a path $Q[v_1,v_2]$ in $Z(v)$ and two
disjoint paths from $I_s(y)$ to $y_1$ and $y_2$, respectively
(Lemma~\ref{2p1to2}). Then there are two disjoint paths from $v_1$
to $I_s(y)$, and therefore $y\in C_s(v)$.
\qed\end{pf}

If $z_x=I_t(z_y)$ and $z_y=I_s(z_x)$ as well as $\beta_1(x,y)$ is
true, we say that $Z(x)$ and $Z(y)$ are a {\em candidate component
pair}. By Lemma~\ref{type2b}, to find an optimal pair of type II, we
only need to determine if there exists a valid pair for any
candidate component pair. For a candidate component pair $Z(x)$ and
$Z(y)$, let $H_{xy}$ be the digraph with vertex set $U=\{v|y\prec
v\wedge v\prec x\}\cup Z(y)\cup Z(x)$. The edge set is
$E(D^+[U])-E(Z(x))-E(Z(y))$, in which $D^+[U]$ is the subgraph of
$D^+$ induced by $U$.

\begin{figure}[t]
\begin{center}
\includegraphics[scale=1.2]{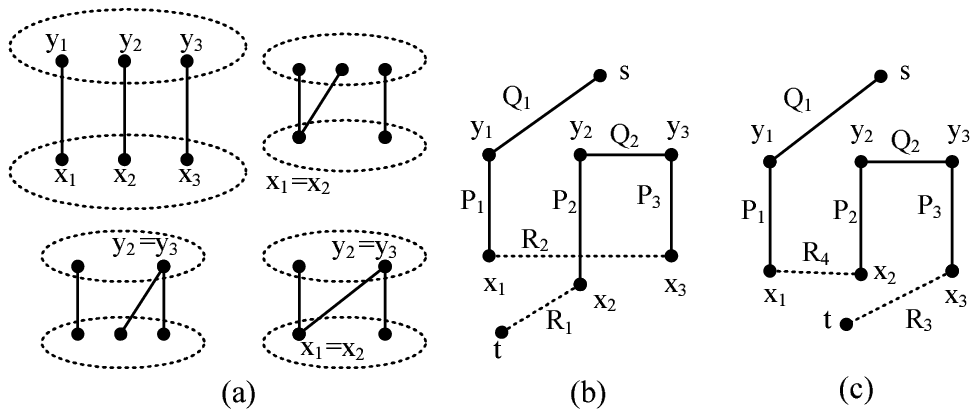}
\caption{{\rm (a)} Four cases of the three paths when $\beta_2(x,y)$ is true; {\rm (b)} The case that $R_1$ and $R_2$ exist ({\rm Lemma~\ref{type2iff}}); {\rm (c)} The case that $R_3$ and $R_4$ exist {\rm (Lemma~\ref{type2iff})}.}
\label{fbeta}
\end{center}
\end{figure}

\begin{defi}
The predicate $\beta_2(x,y)$ is true iff there are three internally
disjoint paths $P_1[y_1,x_1]$, $P_2[y_2,x_2]$ and $P_3[y_3,x_3]$ in
$H_{xy}$ satisfying: (1) $\{x_1,x_2,x_3\}\subset Z(x)$ and
$\{y_1,y_2,y_3\}\subset Z(y)$; and (2) $x_i\neq x_j$ or $y_i\neq
y_j$ for $1\leqslant i,j\leqslant 3$; and (3) $y_1\neq y_2$ and
$x_2\neq x_3$.
\end{defi}

Fig.~\ref{fbeta}.(a) illustrates the four possible cases of the
three paths when $\beta_2(x,y)$ is true.
\begin{lem}\label{3v2path}
For any three vertices $y_1$, $y_2$ and $y_3$ in $Z(y)$, there are
two disjoint paths $P[s,y_1]$ and $Q[y_3,y_2]$, or $P[s,y_2]$ and
$Q[y_3,y_1]$.
\end{lem}
\begin{pf}
Let $p=I_s(y)$ and $R$ be any $sp$-path. Note that $R$ avoids
$Z(y)$. By Lemma~\ref{2p1to2}, there are two disjoint paths
$P_1[p,y_1]$ and $P_2[p,y_2]$. If $y_3$ is on one of the paths, we
have done. Otherwise, let $P_3$ be any path in $Z(y_1)$ from $y_2$
to $y_3$, and $q$ be the last vertex of $P_3$ appeared on $P_1$ or
$P_2$. If $q$ is on $P_1$, then  $Q=P_1[y_1,q]\circ P_3[q,y_3]$ and
$P=R\circ P_2$ are the desired two paths. The case that $q$ is on
$P_2$ can be shown similarly.
\qed\end{pf}
\begin{cor}\label{3v2pathb}
For any three vertices $x_1$, $x_2$ and $x_3$ in $Z(x)$, there are
two disjoint paths $P[x_2,t]$ and $Q[x_1,x_3]$, or $P[x_3,t]$ and
$Q[x_1,x_2]$.
\end{cor}

\begin{lem}\label{type2iff}
Suppose that $Z(x)$ and $Z(y)$ are a candidate component pair. There
exist $x'\in Z(x)$ and $y'\in Z(y)$ such that $(x',y')$ is a valid
backward pair of type II iff $\beta_2(x,y)$ is true.
\end{lem}
\begin{pf}
It is clear that if $(x',y')$ is valid for type II, the three paths
exist and $\beta_2(x,y)$ is true. It remains to prove that such a
valid $(x',y')$ exists if $\beta_2(x,y)$ is true. Since
$\beta_2(x,y)$ is true, there are three internally disjoint paths
$P_1[y_1,x_1]$, $P_2[y_2,x_2]$ and $P_3[y_3,x_3]$ in $H_{xy}$, in
which $y_1\neq y_2$ and $x_2\neq x_3$.

By Lemma~\ref{3v2path} and w.l.o.g., there exist disjoint paths
$Q_1[s,y_1]$ and $Q_2[y_3,y_2]$. In the case that $y_2=y_3$, $Q_2$
contains only one vertex but no edge. If $x_1\neq x_2$, by
Corollary~\ref{3v2pathb}, there exist two disjoint paths
$R_1[x_2,t]$ and $R_2[x_1,x_3]$; or $R_3[x_3,t]$ and $R_4[x_1,x_2]$.
If $R_1$ and $R_2$ exist, the path $Q_1\circ P_1\circ R_2\circ
\bar{P}_3\circ Q_2\circ P_2\circ R_1$ is a desired path, as shown in
Fig.~\ref{fbeta}.(b). That is, $x'=x_1$ and $y'=y_3$. Otherwise
$R_3$ and $R_4$ exist, and the path $Q_1\circ P_1\circ R_4\circ
\bar{P}_2\circ \bar{Q}_2\circ P_3\circ R_3$ is the desired path, as
shown in Fig.~\ref{fbeta}.(c), namely, $x'=x_1$ and $y'=y_2$.

It remains to consider $x_1=x_2$. By the definition of immediate
dominator, there is a path $R$ from $x_3$ to $t$ avoiding $x_1$. The
path $Q_1\circ P_1\circ \bar{P}_2\circ \bar{Q}_2\circ P_3\circ R$ is
a desired path (similar to Fig.~\ref{fbeta}.(c)), in which $x'=x_1$
and $y'=y_2$.
\qed\end{pf}


From $H_{xy}$, we construct a vertex-capacitated digraph $H_{xy}^+$
as follows. First, if any connected component contains exactly one
vertex $u$ in $Z(x)$ and one vertex $v$ in $Z(y)$, we replace the
component by an edge $(v,u)$. Then, we add a new source $y_0$ and a
new sink $x_0$. For each $v\in Z(y)$ there is an edge $(y_0,v)$; and
there is an edge $(v,x_0)$ for each $v\in Z(x)$. The capacity of
vertex $v$ is denoted by $c(v)$. The capacities are assigned as
follows: $c(x_0)=c(y_0)=\infty$; $c(v)=2$ for any $v\in Z(x)\cup
Z(y)$; and $c(v)=1$ for any other vertex. In the following, ``the
max-flow in $H_{xy}^+$'' means the maximum vertex-capacitated flow
from $y_0$ to $x_0$ in $H_{xy}^+$. Note that a vertex-capacitated
digraph can be easily transformed to an edge-capacitated digraph,
and the maximum flow of a vertex-capacitated digraph can be computed
by traditional maximum-flow algorithms.

\begin{figure}[t]
\begin{center}
\includegraphics[scale=1.2]{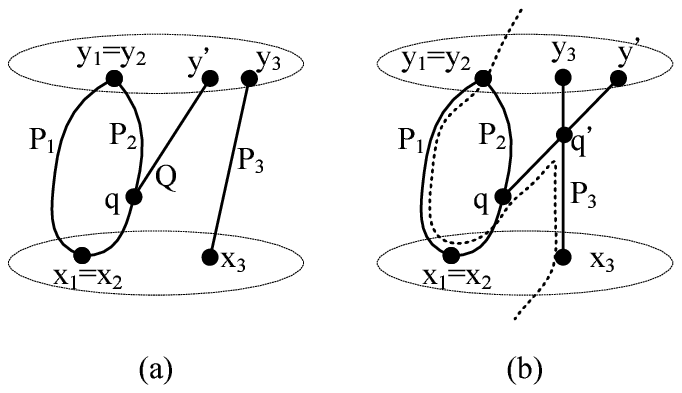}
\caption{Illustrations for Lemma~\ref{beta22}. {\rm (a)} $Q$ and $P_3$ are disjoint; {\rm (b)} $Q$ intersects $P_3$. The dotted-line depicts a valid zigzag path. }
\label{f3back}
\end{center}
\end{figure}
\begin{lem}\label{beta22}
Suppose that $Z(x)$ and $Z(y)$ are a candidate component pair and
$(x',v)$ is not a valid type-I pair for any $x'\in Z(x)$ and
$d_s(v)\geqslant d_s(y)$. Then, $\beta_2(x,y)$ is true iff the
max-flow in $H_{xy}^+$ is at least three.
\end{lem}
\begin{pf}
If $\beta_2(x,y)$ is true, it is easy to see that the max-flow in
$H^+_{xy}$ is at least three. We need to show the other direction.
If there is a flow of value three, there are three internally
disjoint paths $P_i[y_i,x_i]$, $1\leqslant i\leqslant 3$. According
to the assigned capacities, there are at least two distinct vertices
in $\{y_1,y_2,y_3\}$, and so are in $\{x_1,x_2,x_3\}$. The only
question is that two of the three paths may have the same endpoints.
That is, w.l.o.g., $x_1=x_2$ and $y_1=y_2$. By the construction of
$H_{xy}^+$, the connected component containing the two paths must
contain another vertex in $Z(x)\cup Z(y)$. W.l.o.g. let $y'\in Z(y)$
be such a vertex. Then, let $Q$ be a path from $y'$ to $x_1$ and $q$
the first vertex of $Q$ intersecting $P_1$ or $P_2$. W.l.o.g. let
$q$ be on $P_2$ (see Fig.~\ref{f3back}).
\begin{itemize}
\item If $Q$ and $P_3$ are disjoint, the three paths $P_1$, $Q[y',q]\circ P_2[q,x_1]$ and $P_3$ satisfy the requirement and $\beta_2(x,y)$ is true (Fig.~\ref{f3back}.(a)).
\item Otherwise $Q$ and $P_3$ share a common vertex, possibly $y'=y_3$. Let $q'$ be the last vertex of $Q$ on $P_3$.
\begin{itemize}
\item If $q'=y_3$, the three paths $P_1$, $Q[y_3,q]\circ P_2[q,x_1]$ and $P_3$ satisfy the requirement and $\beta_2(x,y)$ is true.
\item Otherwise $d_s(q')\geqslant d_s(y)$.
There exists a path $P_1\circ \bar{P}_2[x_1,q]\circ
\bar{Q}[q,q']\circ P_3[q',x_3]$ is a path from $y_1$ to
$x_3$ with a backward subpath of length $d(q',x_1)\leqslant
d(y,x)$, and this path can be extended to a zigzag path of
type I, a contradiction to the assumption
(Fig.~\ref{f3back}.(b)). Note that $\beta_1(x,q')$ is true
and therefore $d_s(q')<d_s(x)$ by Lemma~\ref{beta1a}.
\end{itemize}
\end{itemize}
\qed\end{pf}

\begin{cor}\label{betatime}
Under the assumption of Lemma~\ref{beta22}, $\beta_2(x,y)$ can be
determined in $O(m_{xy}+n_{xy})$ time, in which $m_{xy}$ and
$n_{xy}$ are the numbers of edges and vertices in $H_{xy}^+$,
respectively.
\end{cor}
\begin{pf}
By Lemma~\ref{beta22}, $\beta_2(x,y)$ can be determined by checking
whether the max-flow in $H_{xy}^+$ is larger than or equal to three.
Since all the capacities are integral, this max-flow question can be
determined with at most three iterations of the augmentation step of
the Ford-Fulkerson maximum flow algorithm \cite{for56,cor01} or
equivalently at most three breadth-first search on the residue
graphs. Therefore the time complexity is linear.
\qed\end{pf}

\begin{lem}\label{type2t}
Suppose that $l$ is the length of an optimal backward subpath of
type I. In linear time, we can find an optimal backward subpath of
type II with length less than $l$ or determine there is no such
subpath.
\end{lem}
\begin{pf}
By Lemmas \ref{type2iff} and \ref{beta22}, we can determine if there
exists an optimal backward subpath of type II with length less than
$l$. Note that the proofs of Lemmas~\ref{type2iff} and \ref{beta22}
are constructive and they implies a linear time algorithm for
constructing such a zigzag path if it exists. By Corollary
\ref{betatime}, the time complexity is linear to
$\sum(m_{xy}+n_{xy})$, in which the summation is taken over all
candidate component pairs. By Lemma~\ref{type2b} and the uniqueness
of immediate dominator, any zero-component is involved in the
max-flow computations at most twice. Therefore the total time
complexity is $O(m+n)$.
\qed\end{pf}

\subsection{Types III and IV}

Types III and IV are similar to Type II, but simpler. Furthermore,
the two types are symmetric and we shall only explain Type III
briefly. A pair $(x,y)$ is valid for type III if $z_y$ is an
$s$-dominator of $z_x$ and $z_x$ is not a $t$-dominator of $z_y$.
\begin{lem}\label{type3}
If $(x,y)$ is an optimal pair of type III, $z_y=I_s(z_x)$.
\end{lem}
\begin{pf}
By using a similar argument as in Lemma~\ref{type2b}, this lemma
follows.
\qed\end{pf}
In the next lemma, $H_{xy}$ has the same definition as in type II.
\begin{lem}\label{type3iff}
Suppose that $z_y=I_s(z_x)$ and $z_x$ is not a $t$-dominator of
$z_y$. There exist $x'\in Z(x)$ and $y'\in Z(y)$ such that $(x',y')$
is a valid backward pair of type III iff there are two disjoint
paths $P_1[y_1,x_1]$ and $P_2[y_2,x_2]$ in $H_{xy}$ such that
$y_1\neq y_2$.
\end{lem}
\begin{pf}
It is clear that if $(x',y')$ is valid for type III, the two paths
exist. Conversely, if $P_1$ and $P_2$ exist, by Lemma~\ref{2p1to2},
there are two disjoint paths $Q_1[I_s(y),y_1]$ and
$Q_2[I_s(y),y_2]$, respectively. Since $z_x$ is not a $t$-dominator
of $z_y$, we can find a path $R$ from $y_1$ to $t$ and avoiding
$Z(x)$. Let $v$ be the last vertex of $R$ intersecting $Q_1$ or
$Q_2$. Then $v$ is valid for $x_1$, namely, $x'=x_1$ and
$y'=v$.
\qed\end{pf}

\begin{lem}\label{type34t}
Suppose that $l$ is the length of an optimal backward subpath of
type I. In linear time, we can find an optimal backward subpath of
type III or IV with length less than $l$ or determine there is no
such subpath.
\end{lem}
\begin{pf}
Similar to Lemma~\ref{beta22}, the necessary and sufficient
condition of type III shown in Lemma~\ref{type3iff} can be
determined by checking whether the max-flow in $H_{xy}^+$ is at
least two or not. And the max-flow computations for all candidate
pairs can be done in linear time. The optimal backward subpath of
type IV can be computed similarly.
\qed\end{pf}

\section{Shortest detour path}

In this section we show an efficient algorithm for finding a
shortest detour path. A shortest detour path contains exactly one
outward subpath and has no backward subpath, in which an outward
subpath is a path $P$ such that $E(P)\subset E-E(D)$, both endpoints
of $P$ are in $V(D)$, and any of its internal vertex is not in
$V(D)$. Note that a \emph{simple} $st$-path containing an edge not
in $D$ must have length strictly larger than $d(s,t)$, or otherwise
it should be entirely in $D$. Our goal is to efficiently find a
minimum length $st$-path with an outward subpath.

In this section $T$ denotes an arbitrary shortest-path tree of $G$
rooted at $s$ and let $F=T-E(D)$ denote the graph obtained by
removing edges in $E(T)\cap E(D)$ from $T$. Apparently $F$ is a
forest consisting of subtrees of $T$ and $V(F)=V(T)=V$. By the
definition of $D$, any shortest path between two vertices in $D$
must be included in $E(D)$. For any $v\in V(D)$, the path from $s$
to $v$ on $T$ must be entirely within $E(D)$ and therefore $v$ must
be a root of a subtree of $F$. Furthermore, the root of any subtree
of $F$ must be in $V(D)$ because the edge between it and its parent
is removed.
\begin{defi}
For any vertex $v\in V$, let $r_v$ denote the root of the subtree of
$F$ which $v$ belongs to. Let $\widetilde{E}$ denote the set of
edges $(x,y)$ such that $(x,y)\in E-(E(T)\cup E(D))$ and $r_x\neq
r_y$.
\end{defi}

Define
\begin{eqnarray}
f(x,y)=\left\{\begin{array}{ll}
d_s(x)+w(x,y)+d_t(y)\;& \mbox{if }(x,y)\in \widetilde{E}\\
\infty&\mbox{otherwise.}
\end{array}
\right.
\end{eqnarray}
Note that, since $G$ is undirected, both $(x,y)$ and $(y,x)$ denote
the same edge. But $f(x,y)\neq f(y,x)$ in general.

\begin{lem}\label{out1}
Any detour path $P$ contains an edge in $\widetilde{E}$.
Furthermore, if $(u,v)\in \widetilde{E}$ is an edge on $P$, then
$f(u,v)\leqslant w(P)$.
\end{lem}
\begin{pf}
By definition $P$ contains an outward subpath. Since the both
endpoints of this outward subpath are in $V(D)$, they must be in
different subtrees of $F$, and $P$ must have an edge in
$\widetilde{E}$. The result $f(u,v)\leqslant w(P)$ directly follows
from definitions.
\qed\end{pf}

For any vertex $v$, $d_t(v)\leqslant d(v,r_v)+d_t(r_v)$ and the
equality holds iff $P[v,r_v]\circ Q[r_v,t]$ is a shortest $vt$-path,
in which $P[v,r_v]$ is the $vr_v$-path in $T$ and $Q[r_v,t]$ is an
arbitrary shortest $r_vt$-path in $D$. A vertex $v$ is a
\emph{dangler} if $d_t(v)=d(v,r_v)+d_t(r_v)$. By definition any
vertex in $V(D)$ is a dangler.
\begin{lem}\label{dangler1}
If $v$ is not a dangler, there exists a detour path $Q$ of length at
most $d_s(v)+d_t(v)$.
\end{lem}
\begin{pf}
Let $P$ be the $sv$-path in $T$ and $P'$ any shortest $vt$-path. Let
$q$ be the last vertex of $P'$ intersecting $P$, possibly $q=v$. The
path $Q=P[s,q]\circ P'[q,t]$ is a simple $st$-path, and the length
of $Q$ is  $d_s(q)+d_t(q)\leqslant d_s(v)+d_t(v)$. We shall show
$d_s(q)>d_s(r_v)$. Then, by the definition of $F$, $r_q=r_v$ and
therefore $q\notin V(D)$. Consequently $P$ is a simple path not
entirely in $D$ and thus a detour path.

Suppose to the contrary that $d_s(q)\leqslant d_s(r_v)$. Since
$d_t(q)+d_s(q)\geqslant d(s,t)= d_s(r_v)+d_t(r_v)$, we have
$d_t(q)\geqslant d_t(r_v)$ and furthermore $d_t(q)-d_s(q)\geqslant
d_t(r_v)-d_s(r_v)$. Since $q$ is on $P'$,
\begin{eqnarray*}
d_t(v)&=&d(v,q)+d_t(q)=(d_s(v)-d_s(q))+d_t(q)\\
&\geqslant& d_s(v)-d_s(r_v)+d_t(r_v)\\
&=&d(v,r_v)+d_t(r_v)
\end{eqnarray*}
which is a contradiction to that $v$ is not a dangler.
\qed\end{pf}

\begin{lem}\label{dangler2}
Suppose that $v$ is not a dangler and $P$ is any shortest $vt$-path.
For any $u\in V(P)$, $d_s(u)>d_s(r_v)$.
\end{lem}
\begin{pf}
Let $u$ be a vertex with $d_s(u)\leqslant d_s(r_v)$. We show that
$P$ cannot contain $u$. Since $r_v$ is on a shortest $sv$-path,
$d(v,u)\geqslant d(v,r_v)$, and therefore $d(v,u)+d_t(u)\geqslant
d(v,r_v)+d_t(u)$. Since $d_s(u)+d_t(u)\geqslant
d(s,t)=d_s(r_v)+d_t(r_v)$, by $d_s(u)\leqslant d_s(r_v)$, we have
$d_t(u)\geqslant d_t(r_v)$. Thus, $d(v,u)+d_t(u)\geqslant
d(v,r_v)+d_t(r_v)$. Since $v$ is not a dangler,
$d(v,r_v)+d_t(r_v)>d_t(v)$, and therefore $u$ is not on any shortest
$vt$-path.
\qed\end{pf}

\begin{figure}[t]
\begin{center}
\includegraphics[scale=1.2]{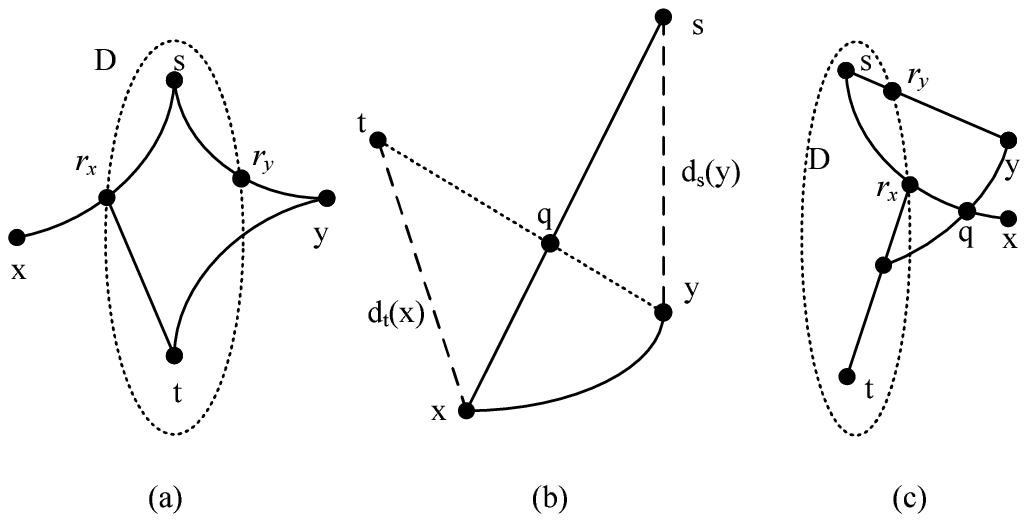}
\caption{{\rm (a)} $x$ is a dangler while $y$ is not. $r_x$ is a common
vertex of a shortest $sx$-path and a shortest $xt$-path; {\rm (b)} The
triangle inequalities when $Y$ intersects $P_x$ (for
Lemma~\ref{outopt}); {\rm (c)} Case 3 in the proof of {\rm Lemma~\ref{outopt}}:
when $q$ is on $P_x[r_x,x]$.  } \label{outf}
\end{center}
\vspace{10pt}
\end{figure}

\begin{lem}\label{outopt}
If $(x,y)$ minimizes function $f$ and $f(x,y)\neq \infty$, then
there exists a simple $st$-path of length $f(x,y)$ and with one edge
in $\widetilde{E}$. Such a path is a shortest detour path.
\end{lem}
\begin{pf}
Since an edge in $\widetilde{E}$ is not an edge in $E(D)$, a simple
path containing edge $(x,y)\in \widetilde{E}$ must have length
strictly larger than $d(s,t)$. We only need to show the existence of
such a simple path, and then it is a shortest detour path by
Lemma~\ref{out1}.

Let $P_x$ and $P_y$ be the shortest paths from $s$ to $x$ and $y$ on
$T$, respectively. Let $Y$ be a shortest $yt$-path. If $P_x$ and $Y$
are disjoint, $P_x\circ (x,y)\circ Y$ is a simple path and its
length is clearly $f(x,y)$. Otherwise let $q$ be the last vertex of
$Y$ intersecting $P_x$. By the triangle inequalities (see
Fig.~\ref{outf}.(b)), we have
\begin{eqnarray*}
f(y,x)&=&d_s(y)+w(y,x)+d_t(x)\\
&\leqslant&w(P_x[s,q]\circ Y[q,y])+w(x,y)+w(P_x[x,q]\circ Y[q,t])\\
&=&w(P_x)+w(x,y)+w(Y)=d_s(x)+w(x,y)+d_t(y)=f(x,y).
\end{eqnarray*}
By the minimality of $f(x,y)$, the equality must hold. i.e.,
\begin{eqnarray}\label{outeq1}
d_s(y)+d_t(x)=d_s(x)+d_t(y).
\end{eqnarray}

We divide into three cases according to whether $x$ and $y$ are
danglers.
\begin{itemize}
\item Case 1: Both $x$ and $y$ are danglers. By Lemma~\ref{2paths}, there are two disjoint paths $Q_1[s,r_y]$ and $Q_1'[r_x,t]$; or $Q_2[s,r_x]$ and $Q_2'[r_y,t]$ in $D^+$.
If $Q_1$ and $Q_1'$ exist, the path $Q_1\circ P_y[r_y,y]\circ
(y,x)\circ P_x[x,r_x]\circ Q_1'$ is a simple path. Since $x$ is
a dangler, $P_x[x,r_x]\circ Q_1'$ is a shortest $xt$-path and
the length of the path is $f(x,y)$. The other case that $Q_2$
and $Q_2'$ exist can be shown similarly.
\item Case 2: Neither $x$ nor $y$ is a dangler. By Lemma~\ref{dangler1}, there exist two detour paths $P_1$ and $P_2$ such that $w(P_1)\leqslant d_s(x)+d_t(x)$ and $w(P_2)\leqslant d_s(y)+d_t(y)$.
Then
\begin{eqnarray*}
\min\{w(P_1),w(P_2)\}&\leqslant& (w(P_1)+w(P_2))/2 \\
&\leqslant& (1/2)(d_s(x)+d_t(x)+d_s(y)+d_t(y))\\
&=&d_s(x)+d_t(y)\; \mbox{(by Eq. (\ref{outeq1}))}\\
&\leqslant& f(x,y).
\end{eqnarray*}
\item Case 3: Either $x$ or $y$ is a dangler. W.l.o.g. assume that $x$ is a dangler but $y$ is not. First we show that $d_s(r_y)<d_s(r_x)$ in this case.
Recall that $Y$ is a shortest $yt$-path intersecting $P_x$. If
the intersection $q$ is on $P_x[s,r_x]$, by
Lemma~\ref{dangler2}, $d_s(r_y)<d_s(q)\leqslant d_s(r_x)$.
Otherwise $q$ is on $P_x[r_x,x]$ and therefore $r_x$ is on a
shortest $yt$-path. As a result, $d(y,r_x)+d_t(r_x)=d_t(y)$.
Since $y$ is not a dangler,
$d(y,r_x)+d_t(r_x)=d_t(y)<d(y,r_y)+d_t(r_y)$. Therefore
\begin{eqnarray}
d_t(r_x)-d_t(r_y)<d(y,r_y)-d(y,r_x). \label{outeq3}\
\end{eqnarray}
Since $r_y$ is on a shortest $sy$-path,
$d_s(r_y)+d(r_y,y)\leqslant d_s(r_x)+d(r_x,y)$, and equivalently
\begin{eqnarray}
d(r_y,y)-d(r_x,y)\leqslant d_s(r_x)-d_s(r_y). \label{outeq2}
\end{eqnarray}
Since both $r_x$ and $r_y$ are in $V(D)$,
$d_s(r_x)+d_t(r_x)=d(s,t)=d_s(r_y)+d_t(r_y)$, and we have
\begin{eqnarray*}
d_s(r_y)-d_s(r_x)&=&d_t(r_x)-d_t(r_y) \\
&<&d(y,r_y)-d(y,r_x)\;\; \mbox{(by Eq. (\ref{outeq3}))}\\
&\leqslant&d_s(r_x)-d_s(r_y).\;\; \mbox{(by Eq. (\ref{outeq2}))}
\end{eqnarray*}
Therefore $d_s(r_y)<d_s(r_x)$. Thus, any shortest $r_xt$-path
$Q$ is disjoint to $P_y$. The path $P_y\circ (y,x)\circ
P_x[x,r_x]\circ Q$ is a simple $st$-path with length
$f(y,x)=f(x,y)$.
\end{itemize}
\qed\end{pf}

\begin{thm}\label{thm:out}
For an undirected graph with nonnegative edge lengths, the shortest
detour path problem can be solved in $O(m+n)$ time if $d_s(v)$ and
$d_t(v)$ are given for all $v$.
\end{thm}
\begin{pf}
By Lemma~\ref{outopt}, the length of a shortest detour path is the
minimum value of function $f$. To compute $(x,y)$ minimizing $f$, we
first construct $D$ and a shortest path tree $T$, and then find the
edge set $\widetilde{E}$. The minimum value of $f$ can be found by
checking both $f(x,y)$ and $f(y,x)$ for all edges $(x,y)\in
\widetilde{E}$. The time complexity is linear if the distances
$d_s(v)$ and $d_t(v)$ for all $v$ are given. Once $(x,y)$ is found,
by the method in the proof of Lemma~\ref{outopt}, the corresponding
path can be constructed in linear time.
\qed\end{pf}

\section{Concluding remarks}
By Theorem~\ref{thm:main}, we have the next corollary.
\begin{cor}
For undirected graphs with nonnegative edge lengths, if the single
source shortest path problem can be solved in $O(t(m,n))$ time, the
next-to-shortest path problem can be solved in $O(t(m,n)+m+n)$ time.
\end{cor}
Important graph classes for which the single source shortest path
problem can be solved in linear time include unweighted graphs (by
BFS \cite{cor01}), planar graphs \cite{hen97}, and integral edge
length graphs \cite{thr99}.

As pointed out in \cite{li06,wu10}, it can be easily shown that the next-to-shortest problem is at least as hard as finding a shortest path between two vertices. When negative-weight edges are allowed, the next-to-shortest problem becomes NP-hard because it is polynomial-time reducible from the longest path problem by a similar reduction. 
An interesting problem is how to efficiently find the next-to-shortest paths for single source and multiple destinations. Another open problem is the complexity of the version on directed graphs with positive-weight edges.

\section*{Acknowledgment}
Bang Ye Wu was supported in part by NSC 97-2221-E-194-064-MY3 and
NSC 98-2221-E-194-027-MY3 from the National Science Council, Taiwan,
R.O.C.

\end{document}